\documentclass[sn-basic]{sn-jnl}

\usepackage{graphicx}%
\usepackage{amsmath,amssymb,amsfonts}%
\usepackage{amsthm}%
\usepackage[title]{appendix}%
\usepackage{listings}%
\usepackage{bm}
\begin{document}

\title[NMF as VAR]{Applying non-negative matrix factorization with covariates to multivariate time series data as a vector autoregression model}

\author[1]{\fnm{Kenichi} \sur{Satoh}}

\email{kenichi-satoh@biwako.shiga-u.ac.jp}
\affil[1]{\orgdiv{Faculty of Data Science}, \orgname{Shiga University}, \orgaddress{\street{Banba 1-1-1}, \city{Hikone}, \postcode{522-8522}, \state{Shiga}, \country{Japan}}}


\abstract{
We propose a novel framework for analyzing multivariate time series (MTS) data by integrating non-negative matrix factorization (NMF) with vector autoregression (VAR). Termed NMF-VAR, this method models the coefficient matrix of NMF as a VAR process, enabling simultaneous extraction of latent components and temporal dependencies. Unlike standard VAR, which struggles with high dimensionality and lacks clarity, our method introduces a low-rank latent structure that reduces the number of parameters while retaining explanatory power.
The proposed framework generalizes the standard VAR model to high-dimensional non-negative data, including the standard VAR as a special case. We formulate the estimation as a constrained optimization problem and present multiplicative update rules for NMF based on existing tri-factorization techniques. 
We evaluate the method on three real-world datasets: quarterly first-differenced macroeconomic indicators of Canada, monthly international airline passenger volumes, and daily COVID-19 infection counts across Japanese prefectures. The results demonstrate that NMF-VAR effectively captures meaningful patterns such as economic cycles, seasonal travel behavior, and regional epidemic trends. Moreover, the method yields a significant reduction in regression parameters, improving both scalability and model transparency. 
Overall, NMF-VAR provides an efficient and insightful tool for analyzing high-dimensional and large-scale time series data.
}


\maketitle

\section{Introduction}\label{sec1}
Non-negative matrix factorization (NMF) has emerged as a powerful tool for dimensionality reduction and feature extraction in high-dimensional non-negative data (\citet{cichocki2009, lee1999, lee2000, gillis2014}). By decomposing a data matrix into the product of two lower-dimensional matrices—the basis and coefficient matrices—NMF provides a clear and interpretable representation of the data.  
A key feature contributing to this clarity is that NMF provides an additive representation, where each observation is approximated by a linear combination of non-negative basis components.

In addition, NMF often produces sparse representations, where many values in the basis and coefficient matrices are zero or close to zero. As a result, each observation is represented using only a few active components. This sparsity not only improves computational efficiency, but also helps to reveal meaningful patterns in the data—such as localized features or distinct parts that make up the whole (\citet{hoyer2004}).

Furthermore, the non-negativity constraints allow the basis and coefficient matrices to be interpreted as probability distributions after appropriate normalization. This probabilistic view enables soft clustering, in which each observation can belong to multiple latent groups with varying degrees of membership.  
This technique has found wide application in various fields, including image analysis, document clustering, and bioinformatics (\citet{xu2003, brunet2004}).

Traditional time series models, such as vector autoregression (VAR) (\citet{lutkepohl2005}) and state-space models (\citet{durbin2012}), are well-suited for capturing temporal dynamics but often struggle with high-dimensional data. To mitigate this issue, various dimension reduction and regularization techniques have been proposed. For example, \citet{stock2002} propose summarizing a large number of predictors using a small number of estimated factors via principal component analysis, forming the foundation of dynamic factor models. More recently, \citet{nicholson2020} introduce hierarchical lag structures that impose structured sparsity on VAR coefficients, enabling both lag selection and improved forecast performance in high-dimensional settings.

In parallel, several studies have explored matrix factorization (MF) frameworks for high-dimensional multivariate time series (MTS). \citet{Yu2016} propose temporal regularized matrix factorization (TRMF), which incorporates an autoregressive structure on the latent embeddings and handles missing values effectively. \citet{Chen2022} further extend this to NoTMF, which accommodates nonstationary time series. Unlike non-negative matrix factorization (NMF), however, these MF-based approaches do not impose non-negativity constraints and thus offer less semantic interpretability.

Although the data types differ, \citet{park2014} used NMF in a latent variable model to analyze categorical time series data.  
\citet{Hooi2019} and \citet{Kawabata2021} focus on online matrices or timestamped series of matrices and propose matrix factorization methods for data observed at each measurement point.  
However, these methods are not natural extensions of VAR for high-dimensional data or for other purposes, as they do not have the structure of a standard VAR, i.e., explaining the observed values with past observations.

To bridge this gap, we propose a framework that integrates NMF with a VAR model to effectively capture both the latent structure and temporal dependencies in MTS data. By representing the coefficient matrix in the NMF decomposition as a VAR model, we can leverage the interpretability of NMF while incorporating the dynamic characteristics of time series data. The proposed model formally includes standard autoregression in special cases, which is a natural extension while still accommodating high-dimensional data.  
This approach is inspired by recent research on incorporating covariates into NMF to improve predictive accuracy and model interpretability (\citet{satoh2023}; \citet{satoh2024}).

To implement the proposed framework, we model each time series as a combination of three components: (1) a latent factor (basis) matrix, (2) a parameter matrix, and (3) a covariate matrix derived from past observations. This structure extends VAR to a latent factor setting, allowing for structured feature extraction and improved explanatory insight.

The rest of this paper is organized as follows.  
Section 2 introduces the proposed NMF-VAR model, which integrates non-negative matrix factorization and vector autoregression by modeling the coefficient matrix as a VAR process.  
Section 3 describes the optimization procedure for estimating the model parameters under non-negativity constraints, using multiplicative update rules adapted from tri-NMF techniques.  
Section 4 presents empirical evaluations on three real-world datasets, demonstrating the proposed method's effectiveness and meaningfulness of the extracted components.
Section 5 discusses the relationship between the proposed framework and related approaches, and provides a comparative analysis.  
The paper ends with a summary and concluding remarks.

\section{NMF-VAR method}\label{sec2}

The proposed method, NMF-VAR, is developed through three steps:
(1) approximating observations via latent factor regression, (2) modeling coefficient vectors using past observations, and (3) reformulating the model as a vector autoregression.
This formulation naturally leads to an extended version of NMF that incorporates temporal structure.

Consider $T$ observations, $\bm{y}_1,\ldots,\bm{y}_T$ of $P$ variables observed at equally spaced time points. We aim to explain the observation $\bm{y}_t$ at time $t$ via past observations.

First, for the observation $\bm{y}_t$, we use a basis matrix $X$ with $Q$ latent factors,
i.e. the latent factor matrix and a coefficient vector $\bm{b}_t$ to make the following approximation:

\begin{equation}
\mathop{\bm{y}_t}_{P \times 1} \approx \mathop{X}_{P \times Q} \mathop{\bm{b}_t}_{Q \times 1}, \qquad t = 1, \dots , T,
\label{yt}
\end{equation}
where $Q$ is assumed to be at most $P$.
The observed value vector is therefore dimensionally reduced by the latent factor matrix to a less dimensional coefficient vector.
In other words, the observed value at time $t$ can be approximated by the sum of constant multiples of the $Q$ latent factors.

Next, we explain the coefficient vector via a parameter matrix $\Theta$ and $D$ past observations, 
$\bm{y}_{t-1},\ldots,\bm{y}_{t-D}$ from $t$:
\begin{equation}
\bm{b}_t = \mathop{\Theta}_{Q \times (PD+1)} 
\begin{pmatrix}
\bm{y}_{t-1}\\
 \vdots  \\
\bm{y}_{t-D}\\
1\\
\end{pmatrix}
=\sum_{d=1}^D \Theta_d \bm{y}_{t-d}+\bm{\theta},
 \mbox{ where }
\Theta=(\mathop{\Theta_1}_{Q \times P},\ldots,\mathop{\Theta_D}_{Q \times P},\mathop{\bm{\theta}}_{Q\times1}).
\label{bt}
\end{equation}
Note that $\bm{\theta}$ corresponds to the intercept.

Then, we provide a brief explanation of the difference in the coefficient vector at time $t$, we have:
\begin{equation}
\bm{b}_t-\bm{b}_{t-1}=\sum_{d=1}^D \Theta_d (\bm{y}_{t-d}-\bm{y}_{t-d-1}),
\label{diffbt}
\end{equation}
By going back in time, we obtain:
\begin{equation}
\bm{b}_t=\bm{b}_{D+1}+\sum_{d=1}^D \Theta_d (\bm{y}_{t-d}-\bm{y}_{D+1-d}).
\label{btpast}
\end{equation}
Thus, the coefficient vectors are also related to the past coefficient vectors.

In summary, the observation at time $t$ is approximated as
\begin{equation}
\bm{y}_t \approx \sum_{d=1}^D \Xi_d \bm{y}_{t-d}+\bm{\xi},
\label{var}
\end{equation}
where $\Xi_d=X\Theta_d$ and $\bm{\xi}=X\bm{\theta}$.
This expresses the VAR coefficients in a low-dimensional form via the basis matrix $X$.
When $X$ is the identity matrix, the model reduces to a standard VAR.

In addition, the proposed method substantially reduces the number of regression parameters. 
While standard vector autoregression (VAR) requires $P(PD + 1)$ parameters, our formulation reduces this to $PQ + Q(PD + 1)$ by introducing a low-rank latent factor structure. 
The relative reduction is quantified by
\begin{equation}
\frac{Q(P + PD + 1)}{P(PD + 1)},
\label{nparam}
\end{equation}
which becomes more pronounced when $Q \ll P$ and $D$ is moderately large. 
This highlights the efficiency and scalability of the proposed approach, especially in high-dimensional time series analysis.

The stability and dynamic behavior of the model (\ref{var}) can be analyzed by reformulating it in companion form.
Define the stacked vector
\begin{equation}
\mathop{\bm{z}_t}_{PD \times 1}=
(\bm{y}_t',\bm{y}_{t-1}',\ldots, \bm{y}_{t-D+1}')',
\label{defz}
\end{equation}
where $\bm{y}_t'$ denotes the transpose of the column vector $\bm{y}_t$, forming a row vector. The vector $\bm{z}_t$ is then constructed by vertically concatenating the current and past $D-1$ observations.
Then, the model can be written as
\begin{equation}
\bm{z}_t \approx F  \bm{z}_{t-1} + \bm{f},
\label{zt}
\end{equation}
where $F$ is the companion matrix and $\bm{f}$ is the intercept vector, defined as
\begin{equation}
\mathop{F}_{PD \times PD} =
\begin{pmatrix}
\Xi_1 & \Xi_2 & \cdots & \Xi_{D-1} & \Xi_D \\
I_P & 0 & \cdots & 0 & 0 \\
0 & I_P & \cdots & 0 & 0 \\
\vdots & \vdots & \ddots & \vdots & \vdots \\
0 & 0 & \cdots & I_P & 0
\end{pmatrix},
\qquad
\bm{f} =
\begin{pmatrix}
\bm{\xi} \\
0 \\
\vdots \\
0
\end{pmatrix}.
\label{Ff}
\end{equation}

This formulation provides a compact representation of the VAR process and facilitates the assessment of model stability by examining the spectral radius of the matrix $F$. This representation is regarded as stable if the spectral radius of $F$, that is, the largest absolute value among its eigenvalues, is strictly less than one. This condition ensures that the influence of past observations gradually diminishes over time and the process does not diverge. 

Moreover, since $\Xi_d = X \Theta_d$, each block in the companion matrix is expressed as a product of low-rank and nonnegative matrices. This structure not only reduces the number of parameters but also limits the degrees of freedom in the system. As a result, it tends to suppress extreme or unstable dynamics, making the process more likely to satisfy the stability condition. This theoretical property is also supported by empirical results presented in Section~4.3, where the relationship between the basis rank $Q$ and the stationarity of the estimated process is examined. This contributes to efficient and stable modeling of high-dimensional time series data in a more interpretable form.

Finally, we show that the proposed model can be equivalently formulated as an extension of NMF that incorporates covariate information.
Given the non-negative observation matrix and covariate matrix as:
\begin{equation}
\mathop{Y}_{P\times (T-D)}=(\bm{y}_{D+1},\ldots,\bm{y}_T), \qquad
\mathop{A}_{(PD+1) \times(N-D)}=
\begin{pmatrix}
\bm{y}_D & \cdots & \bm{y}_{T-1}\\
\vdots & \cdots & \vdots  \\
\bm{y}_1 & \cdots & \bm{y}_{T-D}\\
1 & \cdots & 1\\
\end{pmatrix}
\label{YA}
\end{equation}

Then, we apply non-negative matrix factorization with the number of bases $Q\le \min(P, T-D)$ to obtain the non-negative basis matrix $X$ and parameter matrix $\Theta$:
\begin{equation}
\mathop{Y} \approx \mathop{X}_{P \times Q}B=X \mathop{\Theta}_{Q \times (PD+1)} A
\label{XCA}
\end{equation}
where $B=(\bm{b}_{D+1},\ldots,\bm{b}_{T})$ is called the coefficient matrix.
Even if it is not explained by past observations, an approximation: $Y \approx X B$ can be obtained by the standard non-negative matrix factorization.

Since the coefficient matrix is non-negative, each observation (e.g., at time $t$) can be represented by the coefficient vector $\bm{b}_t = (b_{1,t}, \ldots, b_{Q,t})'$, and the relative contribution of the $q$th basis can be evaluated as a ratio:
\begin{equation}
\frac{b_{q,t}}{\sum_{q'=1}^Q b_{q',t}}, \qquad t = D+1, \ldots, T,
\label{soft}
\end{equation}
which can be interpreted as the membership probability in soft clustering. Based on this probability, each observation can be softly assigned to multiple bases, or alternatively, hard clustering can be applied by assigning it to the basis with the highest membership.  
This enables clustering of observations over the time axis and characterization of temporal patterns.

To enhance the identifiability of the decomposition $Y \approx XB$ in (\ref{XCA}), we impose the constraint that the column sums of the basis matrix $X$ equal one. This normalization eliminates scale ambiguity and allows each column of $X$ to be interpreted as a probability distribution over variables, reflecting the composition of each latent basis.
From a row-wise perspective, the normalized values in each row of $X$ indicate the relative contribution of a variable to the different bases. These values can be used for soft assignment of variables to multiple bases, or for hard clustering by selecting the basis with the highest contribution.
As a result, both rows (variables) and columns (observations) of the data matrix can be clustered probabilistically, providing interpretable groupings from both perspectives.

The decomposition in Equation~(\ref{XCA}) is formally equivalent to the tri-factorization structure proposed in tri-NMF by \citet{ding2006}, which represents the data matrix as the product of three matrices. However, unlike tri-NMF, where all three matrices are treated as unknown, \citet{satoh2023, satoh2024} proposed a formulation in which one of the matrices—specifically, the covariate matrix—is assumed to be known and is not subject to optimization. This allows for prediction based on the covariate matrix \( A \) in Equation~(\ref{XCA}). This formulation is referred to as non-negative matrix factorization with covariates (NMF with covariates). Furthermore, \citet{satoh2024} also considered the case in which the basis matrix is given in advance. In this setting, the model structure becomes equivalent to the mean structure of the growth curve model (GCM), originally proposed by \citet{potthoff1964}.  
To summarize, tri-NMF assumes that all three matrices are unknown; if the covariate matrix \( A \) is known, the model becomes NMF with covariates, and if the basis matrix \( X \) is also known, the model corresponds to the GCM.

\section{Optimization of NMF with Covariates}\label{sec3}

We consider the minimization of the squared Euclidean distance between the observation matrix $Y$ and the approximation matrix $\hat{Y}=X\Theta A$: 
\begin{equation}
D_{EU}(Y,\hat{Y})=\mbox{tr}\{(Y-\hat{Y})'(Y-\hat{Y})\},
\label{eu}
\end{equation}
where $\Theta'$ denotes the transposed matrix of $\Theta$.

When the coefficient matrices $B=\Theta A$ and $\hat{Y}$ are given, the update formula for the basis matrix $X$ remains the same as in standard NMF without covariates.
Using the Hadamard product $\odot$ and Hadamard division $\oslash$, which correspond to the element-wise multiplication and division of matrices, respectively, the update formula can be expressed as
\begin{equation}
X \longleftarrow X \odot (YB' \oslash \hat{Y}B').
\label{renewx}
\end{equation}
The basis matrix is rescaled after each update so that the column sums equal one, which also facilitates later interpretation and comparison.

Next, given the basis matrix $X$ and the current approximation $\hat{Y}$, the update formula for the parameter matrix $\Theta$ is given by:
\begin{equation}
\Theta  \longleftarrow \Theta \odot \{(X'YA' )\oslash (X'\hat{Y}A')\}.
\label{renewc}
\end{equation}

The update formulas discussed above correspond to the case where the covariate matrix \( A \) is treated as known in the tri-NMF framework with a squared Euclidean distance, as studied by \citet{ding2006} and \citet{copar2017}. For the numerical computation of NMF with covariates, the \verb|nmfkc| package\footnote{\url{https://github.com/ksatohds/nmfkc}} for the statistical software R (see \citet{r2024}) developed by \citet{satoh2024} is available.

To ensure convergence, we adopt multiplicative update rules derived from the same principle as \citet{lee1999} and \citet{lee2000}, where the objective function is guaranteed to decrease monotonically through the use of an auxiliary function framework. 
This ensures that repeated updates converge to a locally optimal solution and contribute to the numerical stability of the NMF-VAR method.
In addition, \citet{ding2006} provide rigorous convergence proofs for similar update rules in both two-factor and tri-factor NMF settings, confirming that these algorithms generate a non-increasing sequence of objective values under mild assumptions.

Beyond convergence guarantees, initialization plays a crucial role in the practical performance of the algorithm.
Nonetheless, as \citet{gillis2023} note, multiplicative update algorithms can be sensitive to initialization. To mitigate this, we initialize the basis matrix $X$ using the centroids obtained from K-means clustering of the observation vectors. This approach is based on the fact that each observation can be expressed as a linear combination of the basis vectors. As demonstrated by \citet{fathi2023}, K-means-based initialization enhances both interpretability and convergence performance in NMF algorithms. Moreover, recent advances such as \citet{xiao2024} have also addressed the sensitivity of K-means to initial values.

To enhance interpretability and ensure meaningful factorization, identifiability must also be considered.
Although Lee and Seung (2000) do not explicitly use the term ``identifiability'', they point out that NMF solutions are not unique due to scale and permutation ambiguities in the factor matrices. To address this, we normalize each column of the basis matrix $X$ so that its entries sum to one, making it column-stochastic. This reduces ambiguity and allows geometric interpretation of the factorization. As shown by \citet{gillis2023}, such normalization, combined with mild sparsity, can help achieve partial identifiability. Additional constraints like orthogonality or sparsity, as proposed by \citet{ding2006}, further support identifiability by limiting the solution space.

\section{Examples}\label{sec4}

\subsection{AirPassengers dataset}

The AirPassengers dataset contains monthly totals of international airline passengers from January 1949 to December 1960. This time series dataset consists of observations collected at regular monthly intervals, with each data point representing the number of passengers (in thousands) for a given month. It is included in the \verb|datasets| package in \textsf{R}.

\begin{figure}[h]
\centering
\includegraphics[width=1\linewidth]{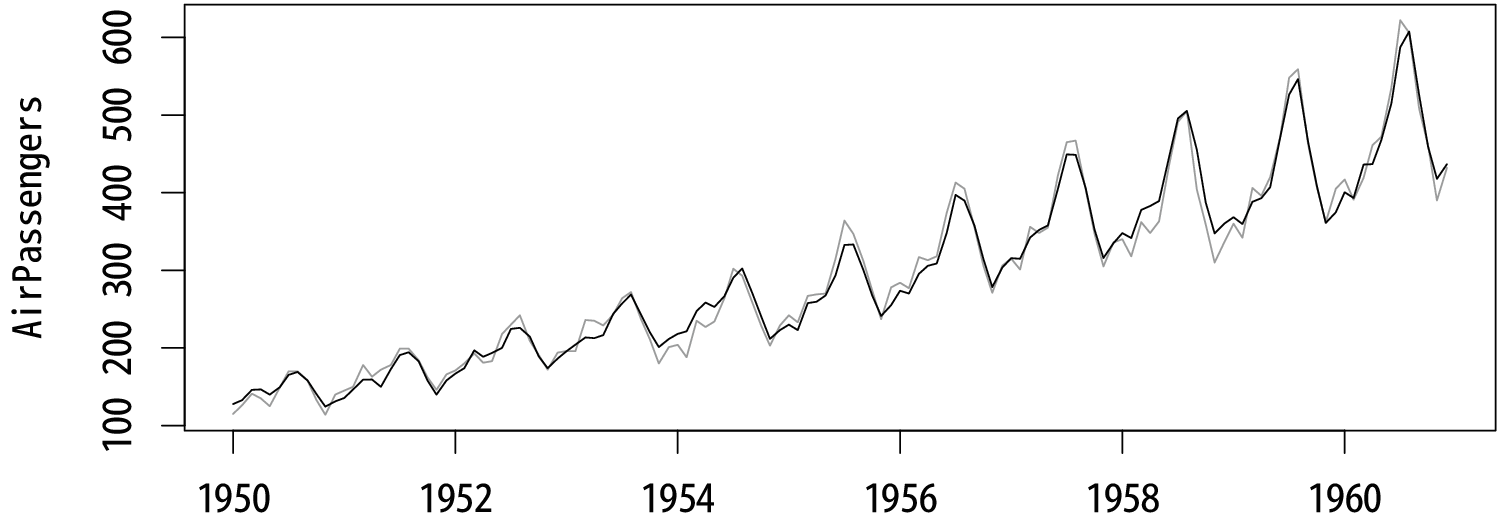}
\caption{Observed values and fitted curves. The solid gray line represents the observed values, and the solid black line represents the fitted values obtained via the NMF-VAR method. The coefficient of determination was $R^2 = 0.982$.}
\label{Fig1}
\end{figure}

Since the observed data are univariate ($P = 1$, $T = 144$), we used a scalar basis $X = 1$, set the number of latent bases to $Q = 1$, and chose a lag order of $D = 12$, using observations from the same month in the previous year as covariates.
After applying a logarithmic transformation to the observations, we applied the NMF-VAR method under these settings. This yielded the following parameter matrix.

\begin{eqnarray}
\Theta=(0.21,
0,0,0,0,0,0, 0,0,0,0.06,0.69,0.11),
\label{c}
\end{eqnarray}
which is rewritten by
\begin{equation}
\bm{y}_t \approx 0.21\bm{y}_{t-1}+ 0.06\bm{y}_{t-11}+0.69\bm{y}_{t-12}+0.11.
\label{d1}
\end{equation}
Although this case resembles a standard autoregressive model, all regression coefficients are constrained to be non-negative due to the use of NMF.
The spectral radius of the companion matrix of (\ref{Ff}) was 0.997, which is slightly below 1, suggesting that the process was just within the boundary of stationarity.
Fig.~\ref{Fig1} shows the observed values and the fitted curves via the NMF-VAR method. 
The coefficient of determination between the observed and fitted values was high, at $R^2 = 0.982$.

\subsection{Canada dataset}

The Canada dataset comprises four key macroeconomic variables representing the Canadian economy.  
These variables are derived from the OECD's Main Economic Indicators, Quarterly National Accounts, and Labour Force Statistics. Specifically:

\begin{itemize}
  \item \verb|e| (employment): Represents the log-transformed number of employed persons, based on civil employment statistics.
  \item \verb|prod| (productivity): A measure of labor productivity, derived from nominal GDP, the consumer price index, and employment data.
  \item \verb|rw| (real wages): Reflects real wages in the manufacturing sector, adjusted using the wage index.
  \item \verb|U| (unemployment rate): The Canadian unemployment rate, expressed as a percentage and taken directly from published statistics.
\end{itemize}

The dataset covers the period from the first quarter of 1980 (1980Q1) to the fourth quarter of 2000 (2000Q4), providing 84 quarterly observations for each variable. Hence, $T = 84$ and $P = 4$.

To address concerns regarding stationarity, we first applied first-order differencing to the multivariate time series $\bm{y} = (\verb|e|, \verb|prod|, \verb|rw|, \verb|U|)'$, and then normalized the differenced series so that the values of each variable ranged within $[0,1]$. The NMF-VAR method was then applied to these normalized differences using $Q = 	2$ basis vectors and a lag order of $D = 1$. The spectral radius of the estimated autoregressive coefficient matrix was $0.778$, indicating that the resulting process is stationary.

The optimized basis and parameter matrices obtained via Equations~(\ref{renewx}, \ref{renewc}) are shown below, with row and column labels included for clarity. Zero entries are indicated by blank spaces.

For compactness, the latent conditions are labeled as ``Cond.1'' and ``Cond.2'' in the matrices. 
These correspond to latent factors capturing distinct macroeconomic states: 
\verb|Cond.1| is primarily composed of \verb|e| and \verb|prod|, and can thus be interpreted as reflecting economic expansion. 
In contrast, \verb|Cond.2| consists mainly of \verb|rw| and \verb|U|, suggesting a recessionary condition.

\begin{eqnarray}
X =
\begin{pmatrix}
& \verb|Cond.1| & \verb|Cond.2| \\
\verb|e|     & 0.55 &        \\
\verb|prod|  & 0.40 & 0.15 \\
\verb|rw|    & 0.06 & 0.41 \\
\verb|U|     &      & 0.44 \\
\end{pmatrix},
\Theta=
\begin{pmatrix}
& \verb|e| & \verb|prod| & \verb|rw| & \verb|U| & const.\\
\verb|Cond.1| & 1.01 & 0.53  &       &      & 0.24\\
\verb|Cond.2| &      &       & 0.54  & 0.88  & 0.20\\
\end{pmatrix}.
\label{XandC}
\end{eqnarray}

\begin{figure}[h]
\centering
\includegraphics[width=1\linewidth]{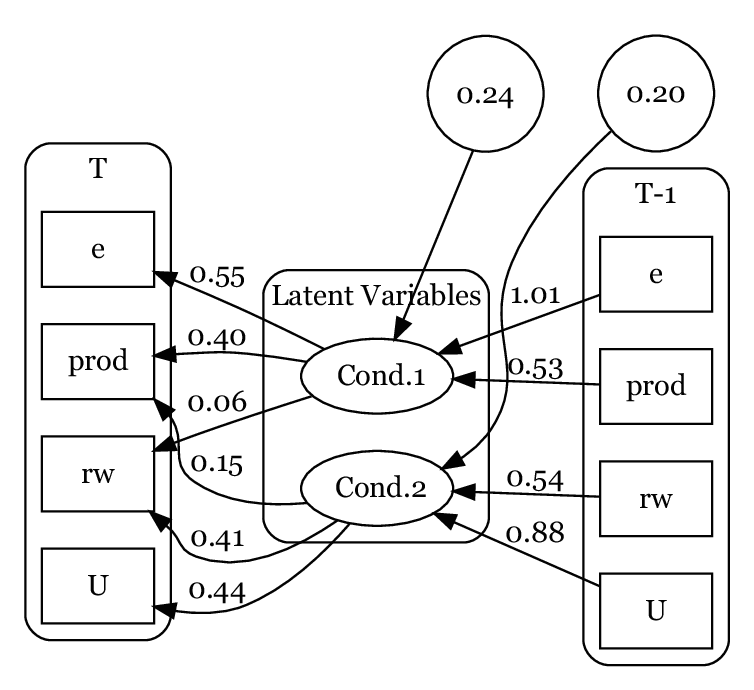}
\caption{
Prediction based on past observations and latent factors.  
Past observations are linearly combined using the coefficients in matrix $\Theta$ to compute the latent coefficients, which are then linearly combined with the latent factors in matrix $X$ to predict the current observations.  
The weights on the arrows correspond to the matrices $X$ and $\Theta$ in Equation~(\ref{XandC}). 
}
\label{Fig2}
\end{figure}

These matrix entries can be interpreted as edge weights in a directed graph structure: past observations influence latent factors via $\Theta$, and these latent factors in turn determine the current observations through $X$, as illustrated in Fig. \ref{Fig2}.

\begin{figure}[h]
\centering
\includegraphics[width=1\linewidth]{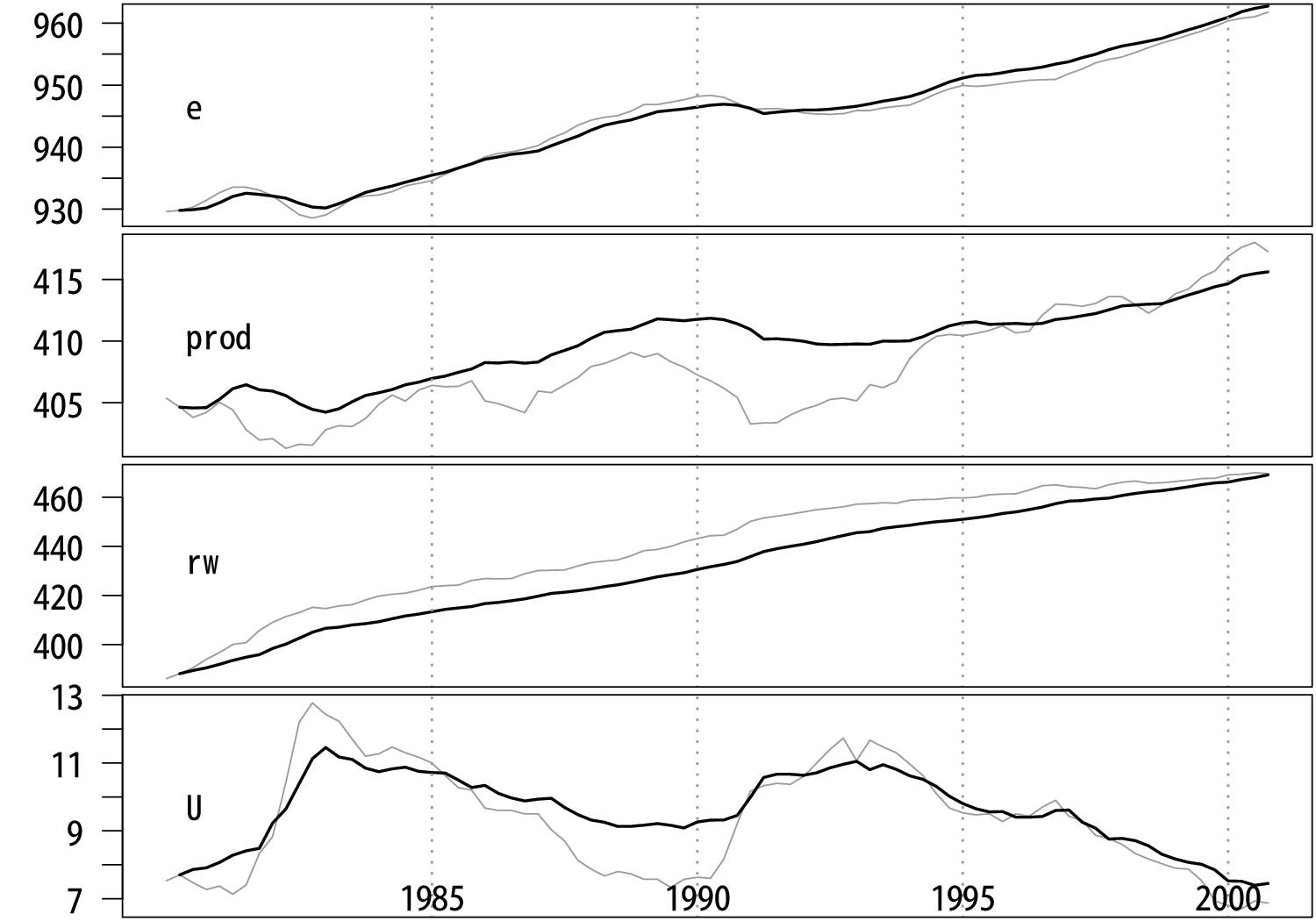}
\caption{
Observed and fitted time series. The solid gray lines represent the observed values, while the solid black lines indicate the fitted values obtained using the NMF-VAR method. The coefficient of determination was $R^2 = 0.599$.
}
\label{Fig3}
\end{figure}

\begin{figure}[h]
\centering
\includegraphics[width=1\linewidth]{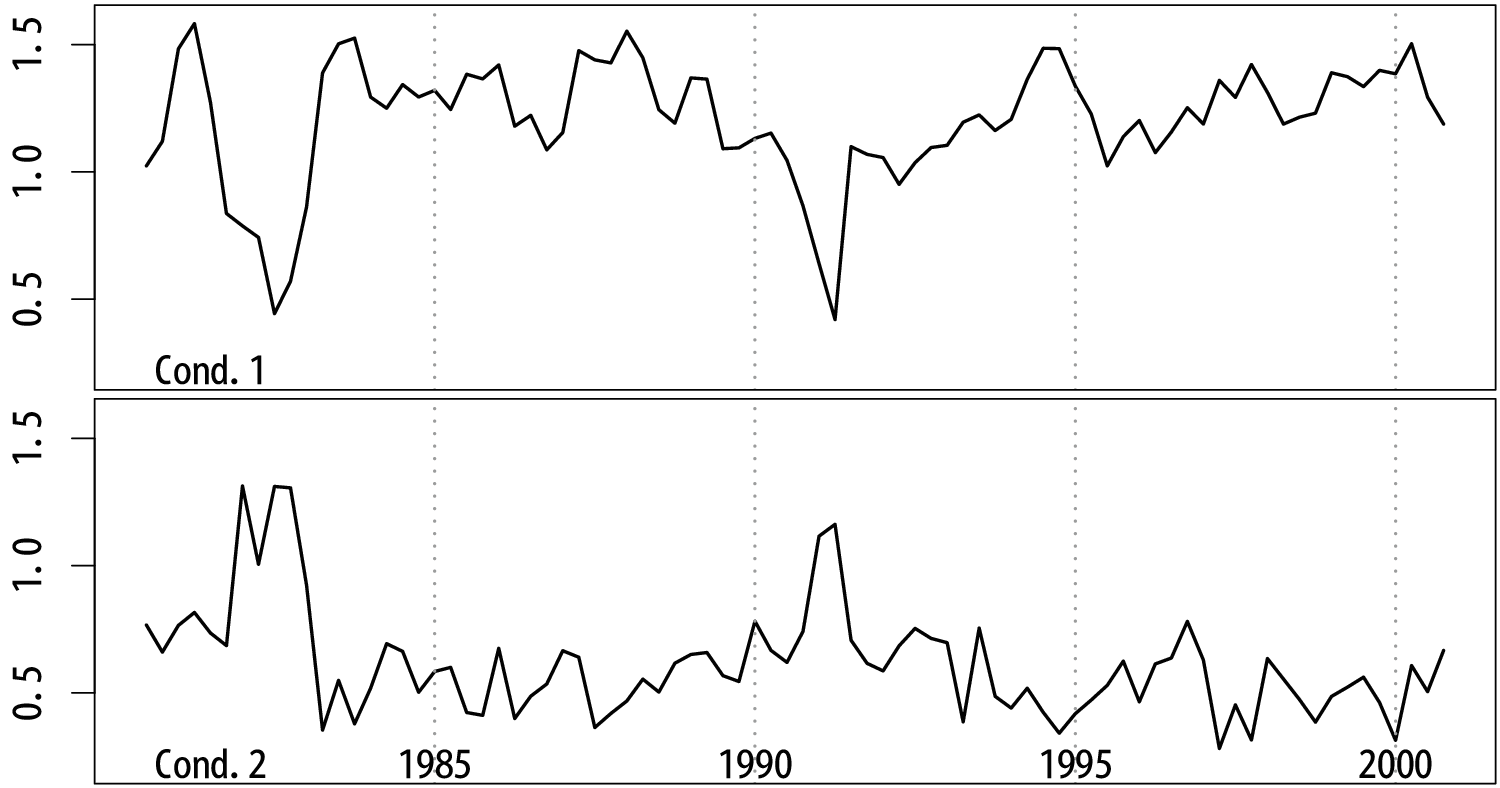}
\caption{
Time-varying coefficients of latent economic conditions estimated by the NMF-VAR method. \texttt{Cond.1} indicates expansion with strong employment and productivity, while \texttt{Cond.2} indicates recession with rising wages and unemployment.
}
\label{Fig4}
\end{figure}

\begin{figure}[h]
\centering
\includegraphics[width=1\linewidth]{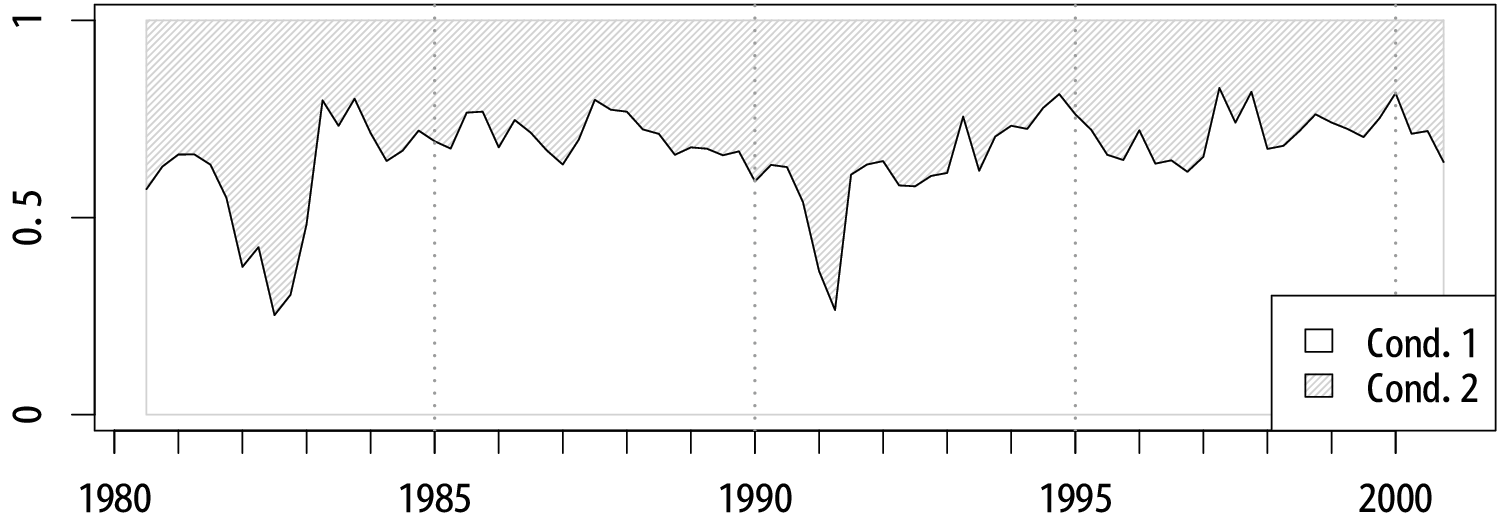}
\caption{
Soft clustering of time points derived from the normalized coefficients in Fig.~\ref{Fig4}.  
Each coefficient vector is scaled to sum to one, yielding the relative contribution (or membership probability) of each latent condition at each time point.  
This representation enables a temporally smooth, probabilistic assignment of states across time.
}
\label{Fig5}
\end{figure}

Fig.~\ref{Fig3} displays the observed values along with the fitted curves obtained using the NMF-VAR method.  
The observed fit, with $R^2 = 0.599$, may be attributed to the complexity of macroeconomic dynamics or to the limited expressiveness of a two-factor model.
Nevertheless, using a small number of latent factors offers the advantage of interpretability, allowing for clearer economic characterization of the underlying components.  
The time-varying coefficient vectors corresponding to the latent factors are presented in Fig.~\ref{Fig4}.  
Their normalized versions, interpreted as membership probabilities, are shown in Fig.~\ref{Fig5}.  
In particular, the membership probabilities in Fig.~\ref{Fig5} provide a single, intuitive indicator that effectively summarizes the state of the economy over time, despite the modest overall fit.

\subsection{COVID-19 dataset in Japan}

As an example of a long-term time series dataset, we treat the daily number of newly infected COVID-19 cases in 47 prefectures in Japan. The data used were prefecture-by-prefecture infection statuses published by the Japan Broadcasting Corporation, NHK, as open data%
\footnote{\url{https://www3.nhk.or.jp/n-data/opendata/coronavirus/nhk_news_covid19_prefectures_daily_data.csv}}.

Observations are available from January 16, 2020 to September 27, 2022, but since daily totals are often zero at the beginning, we use observations from February 13, 2020. To reduce the weekly seasonality arising from delayed reporting over weekends in the COVID-19 data, a 7-day moving average was applied to the daily data. Thus, the observation matrix has $P=47$ and $T=952$ (excluding three days each from the beginning and end—six days in total—due to the 7-day moving average). The observed values were log-transformed by $\log(1+x)$ because the magnitude of the observed values differed depending on the population of each prefecture.

Although a higher number of bases $Q$ tends to increase the coefficient of determination ($R^2$), the interpretability of the bases becomes more difficult. In the analysis of COVID-19 data, interpretation became challenging when $Q=5$, and hence $Q=4$ was adopted. The selection of the number of bases can also be guided by the $IC_p$ criterion proposed by \citet{bai2002}. For the degree of autoregression, 10-fold cross-validation was performed with $Q=4$, and the optimal lag was $D=7$.

When the NMF-VAR method was applied, the coefficient of determination was $R^2 = 0.970$, indicating a good fit to the data. The spectral radius of the companion matrix of (\ref{Ff}) was 0.9945, which is slightly below 1; thus, the estimated VAR process satisfies the condition for weak stationarity. Note that in this case, the number of parameters for the regression coefficients is $PQ + Q(PD + 1) = 1{,}508$, which corresponds to approximately 10\% of the $P(PD + 1) = 15{,}510$ parameters required in standard autoregression. See Equation~(\ref{nparam}) for the reduction rate.

This numerical result complements the theoretical discussion in Section~2.  
To further examine the relationship between the basis rank $Q$ and stationarity, we computed the spectral radius of the companion matrix for various values of $Q$.  
We found that the spectral radius remained below 1 for $Q = 4$ through $Q = 11$, indicating that the estimated VAR process was stationary in these cases.  
For $Q = 12$ through $Q = 19$, the spectral radius consistently exceeded 0.999, and for $Q = 20$ through $Q = 30$, it exceeded 1, indicating that the process became nonstationary.  
These results support the interpretation that a smaller $Q$, by constraining the model’s degrees of freedom, contributes to stability and highlights the stabilizing role of the basis matrix.

\begin{figure}[h]
\centering
\includegraphics[width=1\linewidth]{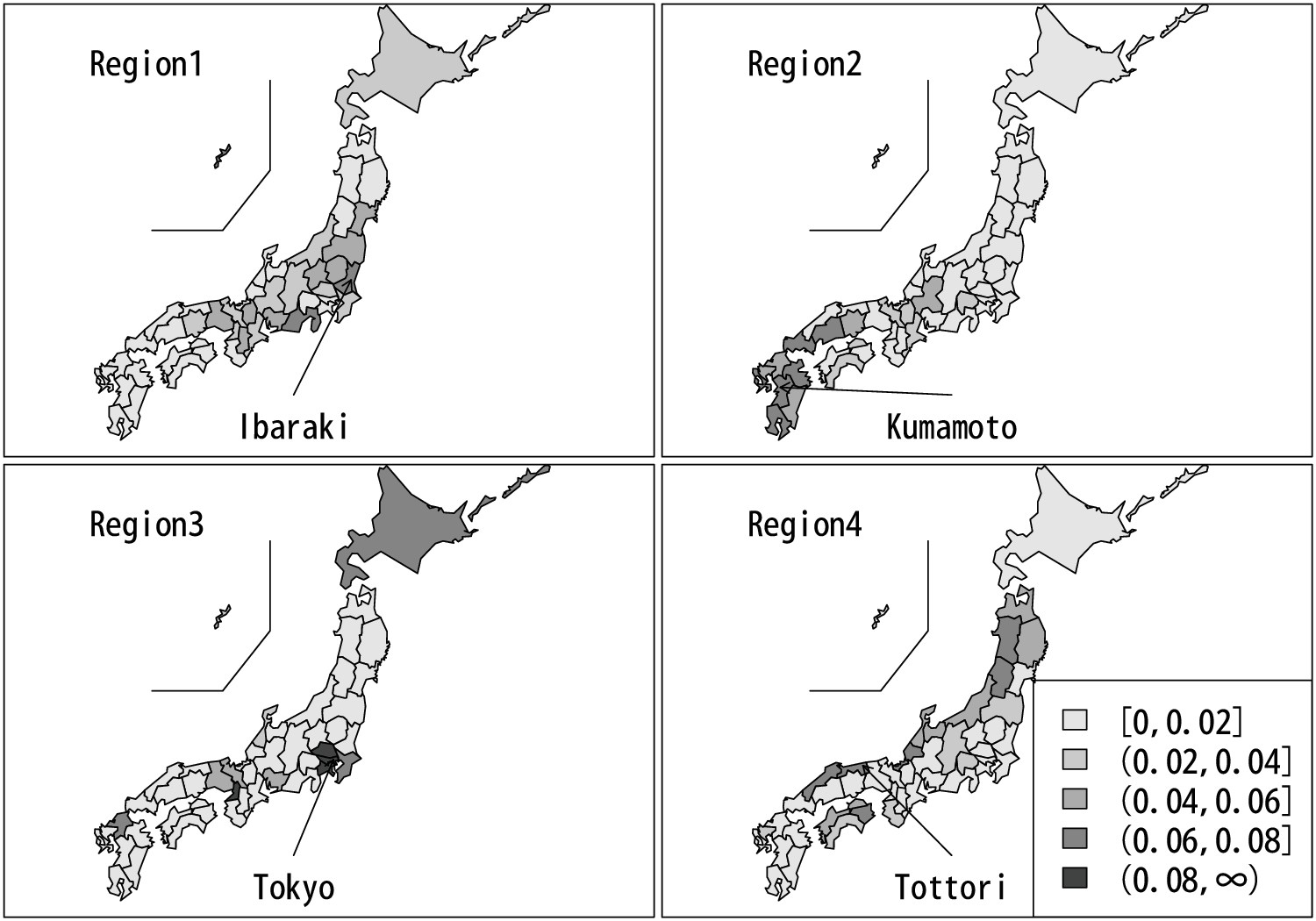}
\caption{
Composition probabilities of prefectures for each latent basis. The basis vectors, referred to as \texttt{Region1} through \texttt{Region4}, represent automatically discovered latent groups based on similar temporal infection patterns. Geographically, \texttt{Region1} corresponds to moderately populous prefectures in the Kanto and Pacific coastal areas, \texttt{Region2} includes mid-sized prefectures in western Japan such as Kyushu and Chugoku, \texttt{Region3} comprises highly urbanized and populous areas including Tokyo, Osaka, Aichi, and Fukuoka, and \texttt{Region4} represents less populated regions along the Sea of Japan, including Tohoku and Hokuriku. Prefectures with high contributions to each basis are labeled on the maps as representative examples. The corresponding time series fits for these prefectures are shown in Fig.~\ref{Fig7}.
}
\label{Fig6}
\end{figure}

The first to fourth columns of the basis matrix correspond to latent factors, labeled \verb|Region1| through \verb|Region4|. Fig.~\ref{Fig6} shows the composition probabilities of prefectures for each latent basis.

In terms of geographical distribution, \verb|Region3| corresponds to highly urbanized and populous areas, including Tokyo, Aichi, Osaka, and Fukuoka. \verb|Region1| includes moderately populous prefectures in the Kanto region and along the Pacific coast. \verb|Region2| covers mid-sized areas in western Japan, including Kyushu and the Chugoku region. \verb|Region4| mainly consists of less populated prefectures along the Sea of Japan, such as those in the Tohoku and Hokuriku regions.

\begin{figure}[h]
\centering
\includegraphics[width=1\linewidth]{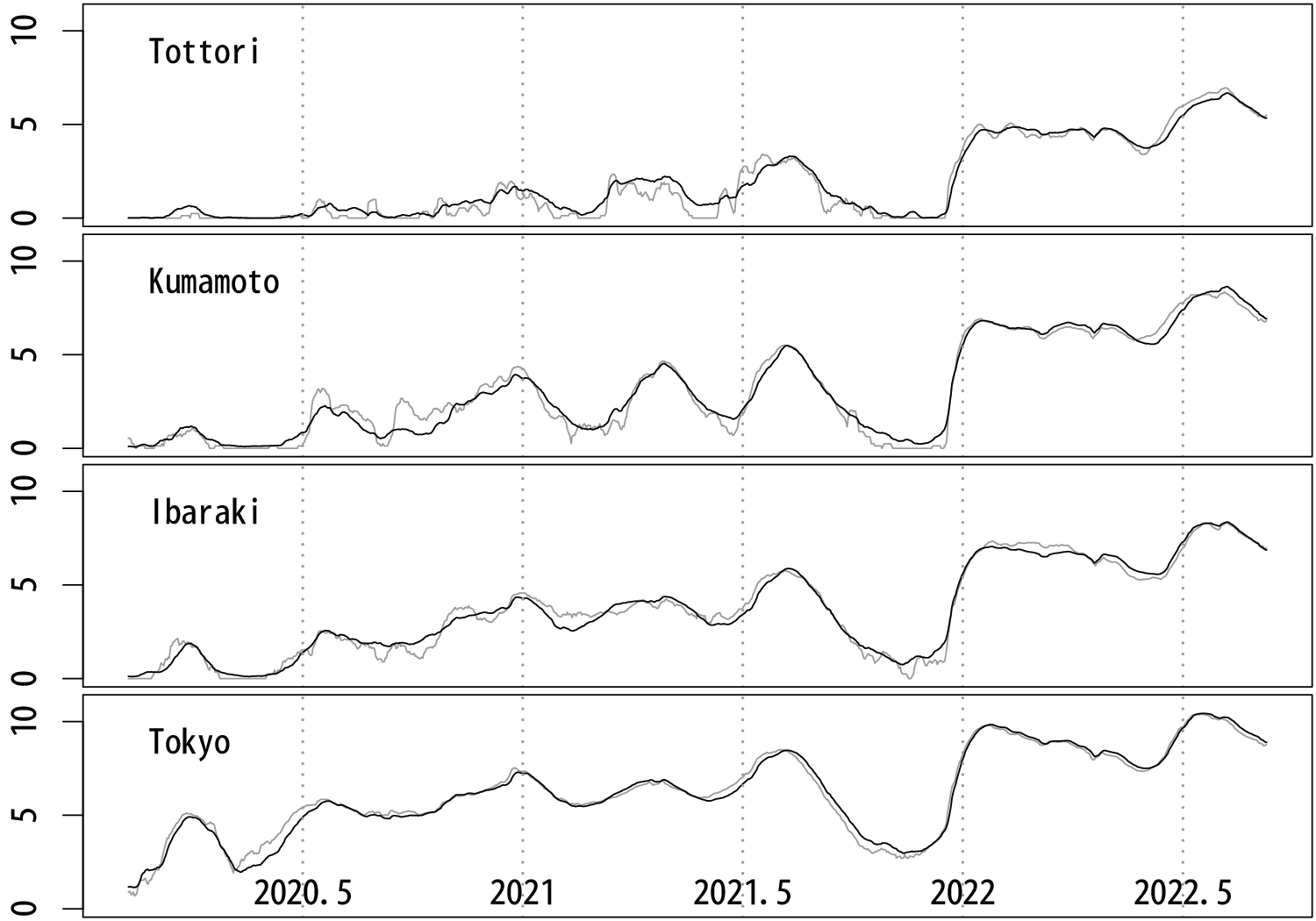}
\caption{
Observed and fitted time series for four representative prefectures, each corresponding to a dominant region in Fig.~\ref{Fig6}. 
The solid gray lines represent the observed values, and the solid black lines represent the fitted values from the NMF-VAR model. 
The coefficient of determination is $R^2 = 0.970$.
}\label{Fig7}
\end{figure}

\begin{figure}[h]
\centering
\includegraphics[width=1\linewidth]{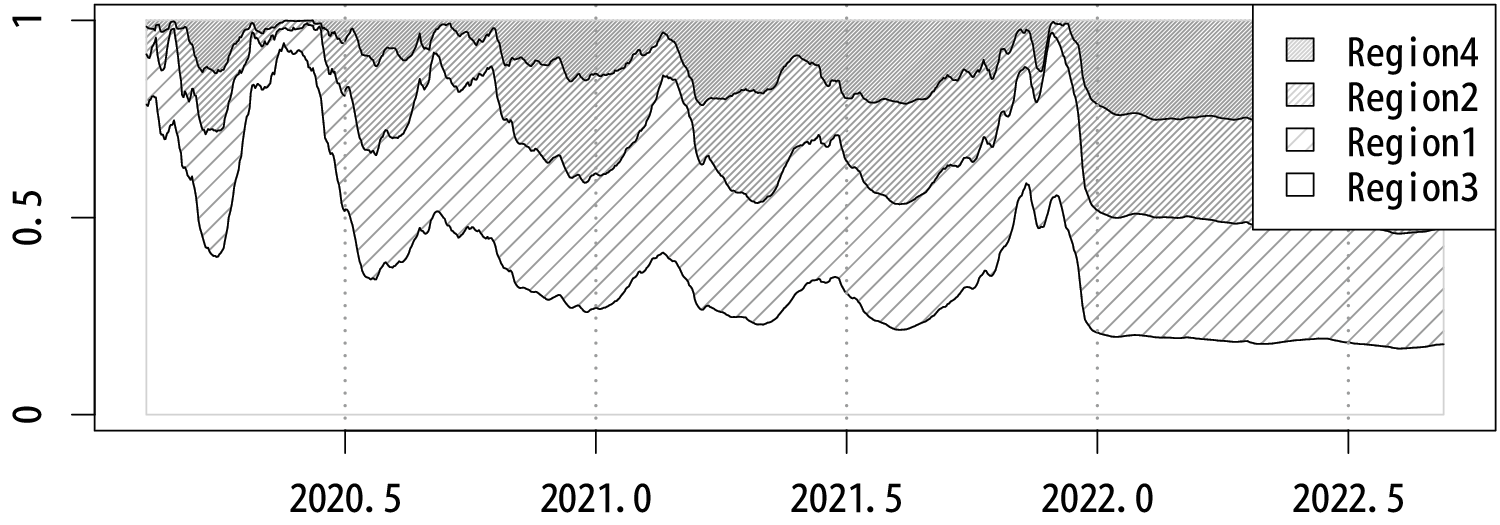}
\caption{
Soft clustering of time trends based on normalized coefficient vectors. 
At each time point, the membership probabilities across all regions sum to one, allowing for a probabilistic interpretation of regional contributions over time. 
This visualization reveals shifts in the dominant geographic patterns of infections throughout the study period.
}\label{Fig8}
\end{figure}

Fig.~\ref{Fig7} and Fig.~\ref{Fig8} provide complementary perspectives on the temporal and geographic dynamics of COVID-19 infections in Japan. 

For instance, the values in the row of the basis matrix corresponding to Ibaraki are $(0.069, 0.002, 0.005, 0.001)$, 
which are normalized to yield contribution probabilities of $(0.90, 0.03, 0.06, 0.02)$ across the four components, 
indicating that Ibaraki is predominantly represented by \verb|Region1|.
Similarly, Kumamoto is best explained by \verb|Region2|, and Tokyo and Tottori have their highest membership in \verb|Region3| and \verb|Region4|, respectively. 
Fig.~\ref{Fig7} shows the observed and fitted time series for these four representative prefectures, each corresponding to a dominant region.

Fig.~\ref{Fig8} illustrates the soft clustering of time trends, where each time point is assigned probabilistic membership to multiple regions. 
While this temporal progression may resemble a spread of infections across regions, the underlying dynamics could also reflect varying local epidemic intensities occurring in parallel.
This framework captures not only temporal changes in regional dominance, but also allows for continuous (rather than discrete) transitions in spatial contribution patterns.
In early 2020, \verb|Region3|—which includes highly urbanized and populous areas—dominates the composition. 
From the second half of 2020, \verb|Region1| becomes more prominent, and in 2021, \verb|Region2| and \verb|Region4| begin to increase in their contributions. 
This pattern suggests a gradual geographic shift in the dominant regions over time.

To examine whether the apparent temporal shifts in Fig.~\ref{Fig8} reflect lagged inter-regional dynamics, we computed the correlation coefficients between the fitted values of Tokyo and those of Ibaraki, Kumamoto, and Tottori, allowing for time lags of up to seven days in either direction. In all cases, the highest correlation was observed at zero lag, suggesting that epidemic trends in these regions occurred largely simultaneously. Although Fig.~\ref{Fig8} suggests a staggered rise in regional contributions, this likely reflects differences in relative magnitude rather than sequential transmission. The underlying temporal dynamics of infections appear broadly synchronous across regions, despite variations in their geographic and temporal prominence.

It should be noted, however, that this analysis was limited to four representative prefectures, and a more comprehensive investigation involving all prefectures would be needed to fully characterize the spatiotemporal spread of the epidemic. One possible extension would be to apply the proposed method to time-lagged data in order to explore asynchronous transmission patterns more systematically. However, this would require adjusting for prefecture-specific time shifts prior to factorization. 
Since the number of possible combinations is extremely large, practical implementation would require careful methodological consideration and further development.

\section{Discussion}\label{sec5}

\begin{figure}[htbp]
\centering
\includegraphics[width=1\linewidth]{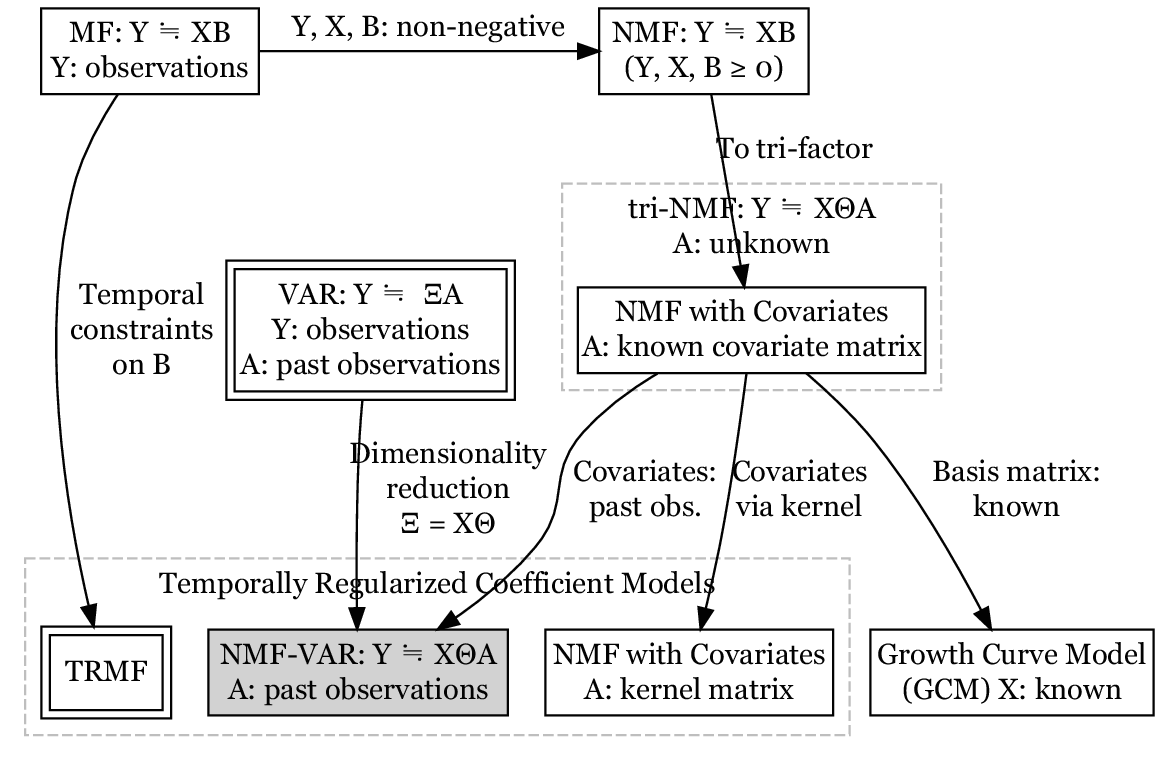}
\caption{Conceptual relationships among matrix factorization models. The proposed NMF-VAR is positioned as an extension of NMF with covariates that integrates a VAR structure using past observations.}
\label{Fig9}
\end{figure}

In this section, we review related matrix factorization methods and position the proposed NMF-VAR model within a broader methodological context.  
Fig.~\ref{Fig9} illustrates the conceptual relationships among the matrix factorization models considered in this study. Standard NMF decomposes the observation matrix into a product of non-negative basis and coefficient matrices. Tri-factor models, such as tri-NMF and kernel-based NMF, introduce a third parameter matrix and allow for the incorporation of known or kernel-derived covariates. Temporal extensions like TRMF impose temporal constraints on the coefficient matrix. In contrast, the proposed NMF-VAR model explicitly incorporates past observations as covariates, aligning with the VAR framework to achieve both dimensionality reduction and structured temporal modeling.

Conceptually, the proposed NMF-VAR model can be regarded as a special case of non-negative matrix factorization with covariates, in which the covariates are constructed from lagged observations. In contrast, standard NMF can be directly applied to non-negative time series data without accounting for temporal structure. While such an approach often yields high approximation accuracy, it sacrifices interpretability due to the unstructured nature of the coefficient matrix.

To address this trade-off between interpretability and approximation accuracy, \citet{satoh2023} proposed the use of kernel-based covariates, as defined in Equation~(\ref{Kernel}), which enable temporal smoothing while preserving model fit. Here, we denote the $T$ measurement time points of the observations as $t_1, \ldots, t_T$.

\begin{eqnarray}
\mathop{Y}_{P \times T} \approx X \mathop{\Theta}_{Q \times T} A,\qquad 
\mathop{A}_{T \times T} =
\begin{pmatrix}
K(t_1,t_1) & \cdots & K(t_1,t_T) \\
\vdots & \ddots & \vdots \\
K(t_T,t_1) & \cdots & K(t_T,t_T) \\
\end{pmatrix}.
\label{Kernel}
\end{eqnarray}
where $K(t_i,t_j) = \exp(-\beta|t_i - t_j|^2), \quad i, j \in \{1,\ldots,T\}$ and $\beta > 0$.

\begin{figure}[htbp]
\centering
\includegraphics[width=1\linewidth]{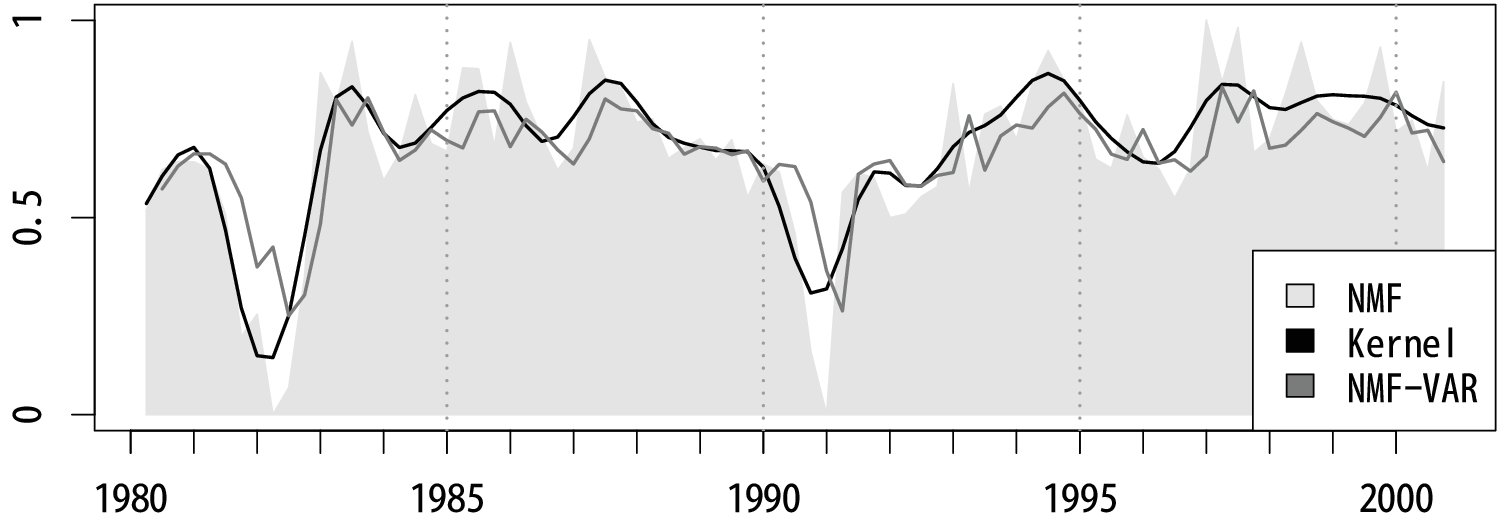}
\caption{Comparison of temporally regularized coefficient models on the Canada dataset. This figure visualizes the soft clustering results from standard NMF, kernel-based NMF, and the proposed NMF-VAR, using the same format as Fig.~\ref{Fig5}. It concretely illustrates the conceptual structure shown in Fig.~\ref{Fig9}, and highlights how temporal regularization affects interpretability and smoothness.}
\label{Fig10}
\end{figure}

To illustrate the concept of temporally regularized coefficient models introduced in Fig.~\ref{Fig9}, we present a comparative visualization in Fig.~\ref{Fig10}. This figure shows the soft clustering results obtained using standard NMF, the kernel-based method, and the proposed NMF-VAR, following the presentation format of Fig.~\ref{Fig5}. The comparison highlights how different modeling strategies for the coefficient matrix affect temporal dynamics.
Standard NMF, which imposes no constraints on the coefficient matrix, exhibits the greatest temporal variability. In contrast, NMF-VAR yields moderately smoothed temporal dynamics by incorporating past observations into the coefficient modeling. The kernel-based method shows the smoothest transitions, achieved through the time-based kernel matrix defined in Equation~(\ref{Kernel}). 
These results demonstrate that introducing temporal structure into the coefficient matrix can effectively suppress excessive variation and enhance interpretability over time.

Although NMF-VAR does not achieve the highest level of goodness-of-fit among the compared methods, it maintains a reasonable degree of fit while offering significant advantages in explanatory clarity. As shown in Table~\ref{Tab1}, the $R^2$ values for NMF-VAR are 0.599 for the Canada dataset and 0.970 for the COVID-19 dataset. 

In particular,  the slightly lower $R^2$ of NMF-VAR compared to standard VAR for the COVID-19 dataset is primarily due to the non-negativity constraint and dimensionality reduction imposed by the low-rank factorization, which favor interpretability and parsimony over pure predictive accuracy.

While the $R^2$ value for NMF-VAR on the Canada dataset is relatively low, this reflects the trade-off involved in achieving improved interpretability through low-rank modeling.
This moderate fit may be expected given the complexity of macroeconomic dynamics and the use of only two latent factors for interpretability.
This structure allows the model to capture temporal dependencies and the influence of past observations on current values, thereby providing meaningful insights into the underlying system dynamics, as demonstrated in the analyses of both the Canada and COVID-19 datasets in Sections~4.2 and 4.3.

\begin{table}[htbp]
\centering
\caption{Coefficient of determination ($R^2$) for various methods applied to the Canada and COVID-19 datasets.}
\label{Tab1}
\begin{tabular}{lccccc}
    \hline
    Dataset & NMF & TRMF & Kernel & NMF-VAR & VAR \\
    \hline
    Canada  & 0.826 & 0.825 & 0.742 & 0.599 & 0.662 \\
    COVID-19 & 0.974 & 0.973 & 0.974 & 0.970 & 0.999 \\
    \hline
\end{tabular}
\end{table}

Finally, we emphasize that the proposed NMF-VAR model is computationally feasible. Although the number of autoregressive coefficients increases with both the number of variables $P$ and lag degree $D$, the effective dimensionality is reduced via the basis matrix. 
As shown in Equation~(\ref{nparam}), the number of parameters required by NMF-VAR is significantly reduced compared to a standard VAR model—particularly when the number of latent bases $Q$ is much smaller than the number of variables $P$ and the lag degree $D$ is moderate.

While this study has focused on temporal and spatio-temporal datasets, future extensions could incorporate spatial adjacency information to reduce model complexity and improve computational efficiency. The NMF-VAR framework, which embeds a VAR structure into the coefficient matrix, can naturally be extended to capture spatial and spatio-temporal dependencies using adjacency-based covariates. Although such extensions may increase the number of parameters, the use of latent variables in the basis matrix enables effective dimensionality reduction. Moreover, the parts-based interpretability inherent in NMF provides valuable insights for analyzing complex, high-dimensional data. These features suggest that NMF-VAR is not only a practical tool for time series analysis but also a promising foundation for future interpretable modeling in spatio-temporal domains. In particular, \citet{wan2023} demonstrates that incorporating spatial distances into regularized VAR estimation effectively reduces model complexity and improves estimation accuracy, highlighting a potential direction for extending the NMF-VAR framework.

\section*{Acknowledgements}

The author thanks the anonymous reviewers for their constructive comments and insightful suggestions, which helped improve the clarity and quality of this paper.

\section*{Statements and Declarations}

\textbf{Funding.} This work was partly supported by JSPS KAKENHI Grant Numbers 22K11930, 25K15229, 	24K03007,  25H00482 and the project research fund by the Research Center for Sustainability and Environment at Shiga University.

\textbf{Conflicts of Interest.} On behalf of all authors, the corresponding author states that there is no conflict of interest.

\textbf{Ethical Approval.} Not applicable.

\textbf{Data Availability.} The datasets generated or analyzed during the current study are available from the corresponding author on reasonable request.

\textbf{Code Availability.} The R package \texttt{nmfkc} is available at \url{https://github.com/ksatohds/nmfkc}.

\bibliography{nmfasvar2.bib}

\end{document}